\documentstyle[11pt]{article}
\begin{document}

\begin{center}
\large \bf
First principles studies of modulated Co/Cu superlattices\\ with strongly 
and weakly exchange biased Co-monolayers
\end{center}
\vspace{1cm}

{\large S.~Krompiewski$^\dagger$, F.~S\"uss* and
 U. Krey*{\footnote{corresponding author, FAX xx49 941 943 4544,
e-mail:
krey@rphs1.physik.uni-regensburg.de}}}
\vspace{1cm}

\normalsize {\it $^\dagger$ Institute of Molecular Physics,
 P.A.N., Smoluchowskiego 17, PL-60-179 Pozna\'n, Poland \\
${}^*$ Institut f\"ur Physik II, Universit\"at Regensburg,D-93040
 Regensburg, Germany}
 \vspace{1cm}

 Received ........... March 1996

superlattice structures). \newline

 \vspace{1cm} \hrule \vspace {0.5cm} {\bf Abstract.}

First-principles calculations have been performed in order to
determine effective exchange integrals between {\it strongly} and {\it
weakly} exchange-coupled Co monolayers in certain modulated periodic
$CoCu_2/CoCu_n$-type superlattices with three non-equivalent Co
planes, which have not yet been studied hitherto. For $3\le n\le 6$ we
find that the two non-equivalent exchange integrals have opposite
signs, i.e.~the strong coupling is antiferromagnetic
 and the weak coupling ferromagnetic, and differ for $n\ne 4$ 
from each other by one order of
magnitude. It is shown that the results depend on the system as a
whole and could not be obtained from separate parts. Finally we suggest
that ''spin valve'' systems of such kind should be considered when
trying to obtain good magneto-resistance together with low
 switching-fields.  \vspace{0.5cm}\hrule \vspace {0.5cm}

Magnetic multilayers based on magnetic transition metals with
nonmagnetic spacers have been intensively studied for almost five
years now, after it was realized that they reveal unusual oscillatory
behaviour of the exchange coupling and magnetoresistance
\cite{l:par1}.  The oscillatory phenomena have a universal character,
do not depend drastically on the kind of metals involved \cite{l:par2}
and occur both with the {\it spacer thickness} as well as {\it
magnetic-slab thickness} variations \cite{l:brun,l:blo1,l:kr1}, and
even depend sensibly on the thickness of an additional non-magnetic
{\it cap-layer} on top of a Co/Cu/Co trilayers system,
e.g.~\cite{l:vries}. The explanation of these phenomena e.g.~by the
essentially
 equivalent quantum confinement,
\cite{l:edwards}, and Fabry-Perrot like electron reflection
theories \cite{l:bruno}, are also by now clear, and not at the center
of our paper. 

However, recently a great deal of attention has been attracted by
exchange-biased ''spin-valve'' systems of the type $AF/F_1/S/F_2$
\cite{l:dien}, with one ferromagnetic slab ($F_1$) {\it strongly}
coupled to an antiferromagnet ($AF$) (e.g. $MnFe$, $CoO$ or $NiO$) and
the other slab ($F_2$) -- almost free -- only {\it weakly} coupled to
the first one via the spacer ($S$).  Systems of this type are not only
interesting for fundamental aspects but may also be applied in future
magnetic recording devices; in particular, in this way one hopes to obtain
  systems, where the spin direction of the
weakly coupled layer is easily flipped (which explains the name ''spin
valves''), while at the same time the resistance of the system is
sensibly changed by the flipping, which is the magneto-resistive effect
mentioned below.

The purpose of the present letter is, to study by reliable {\it
ab-initio}-calculations the question, to which extent it 
would be  possible
to replace the exchange-biasing unit $AF$ in the above-mentioned
conventional ''spin valve system'' by a trilayer
$ferromagnet1/spacer/ferromagnet2$ involving e.g.~only 
ferromagnetic $Co$
planes, and $Cu$ as spacer, provided the thickness of the spacer is
chosen such as to ensure strong antiferromagnetic coupling of the two
ferromagnetic layers \cite{l:blo2}.

In an attempt to get more insight into the nature of exchange coupling
and possible magnetic phases in  such novel systems (see
below), we have studied systematically by the spin-polarized {\it ab
initio LMTO-ASA} method (linearized muffin-tin orbitals, atomic sphere
approximation, in the scalar-relativistic version, see [5]) the series
of modulated periodic  multilayers with supercell
($Co^{(2)}Cu_2Co^{(1)}Cu_2Co^{(2)}Cu_nCo^{(3)}Cu_n$) of the (001)
face-centred tetragonal structure (i.e. $Co$ is grown epitaxially on
$Cu$). These systems are quite specific and have never been studied
before, according to our knowledge: They contain three non-equivalent
$Co$-planes, and as we will see below, their properties cannot simply
be obtained from the behaviour of conventional (1-$Co/$n-$Cu$)$_\infty$
multilayers, which we also have studied for comparison.

 In our modulated systems, the monolayers $Co^{(1)}$ couple strongly
antiferromagnetically with $Co^{(2)}$, while the $Co^{(3)}$ monolayers
turns out to be only weakly coupled (for $n>2$). One of the relevant
questions is the {\it sign} of this coupling (see below).
 The reader should note further that for computational convenience
 our systems are infinite multilayers, i.e.~the above-mentioned
 supercell is periodically continued.

We have built our novel structural models on a similar basis as the
simpler conventional models in our earlier papers \cite{l:kr1,l:kr2},
in particular the in-plane atomic spacings are assumed to be equal to
those of the $fcc-(001)\, Cu$ with the lattice constant $a=3.615$
\AA. Our main task has been to determine both the
 {\it strong} exchange coupling $J$ between $Co^{(1)}$ and $Co^{(2)}$
 as well as the
{\it weak}  coupling $j$ between $Co^{(2)}$ and $Co^{(3)}$ 
from accurate total
energy band calculations for all the relevant spin configurations,
namely: 

\begin{eqnarray*} \begin{array}{lll} 
 (i) & (Co^{(2)}\downarrow Cu_2\,\,
Co^{(1)}\uparrow Cu_2\,\, Co^{(2)}\downarrow Cu_n \,\, Co^{(3)}
\downarrow
 Cu_n)_{\infty} &\quad
([\downarrow\uparrow\downarrow,\downarrow]) \, , \\ (ii) &
(Co^{(2)}\downarrow Cu_2 \,\,Co^{(1)}\uparrow Cu_2 \,\, Co^{(2)}
\downarrow Cu_n \,\,Co^{(3)}\uparrow)
 Cu_n)_{\infty}
 &\quad ([\downarrow\uparrow\downarrow,\uparrow]) \, , \\
(iii) & (Co^{(2)}\uparrow Cu_2 \,\,Co^{(1)}\uparrow Cu_2 \,\,Co^{(2)}
\uparrow Cu_n \,\,Co^{(3)}\uparrow 
Cu_n)_{\infty} &\quad ([\uparrow\uparrow\uparrow,\uparrow]) \, . \\
\end{array}
\end{eqnarray*}

Obviously, since in the present studies no anisotropy is included, all
the systems are spin-rotationally invariant, and there is no
distinction whatsoever between the above mentioned configurations and
the ones with all the spins rotated simultanously by an arbitrary
angle.  After having computed the total energies of the above
configurations ($E_1, E_2$ and $E_3$) the corresponding above-mentioned
$Cu$-mediated weak resp.~strong exchange coupling
integrals have been directly found from 
\begin{eqnarray}\label{eqn1} j =
\frac{1}{4} (E_2-E_1) /A \, , \quad J = -\frac{1}{4} (E_3-E_1) /A \, ,
\end{eqnarray}

where $A$ is the  cross-section area of the unit supercell and $E_i$
are the energies per supercell in the above-mentioned
states. Furthermore,
 one factor of $\frac{1}{2}$ in eqn.~(1) comes from the fact
that there are two thick spacers (related to the weak exchange
coupling $j$) and two thin spacers (related to the large one,
$J$), whereas the other factor of $\frac{1}{2}$ results from
 the spin flip
process according to the well known Heisenberg interaction energy per
''bond'' $<ij>$:

\begin{equation} E_{<ij>} = -J_{ij}
\frac{\vec{S}_i\cdot\vec{S}_j}{\mid \vec{S}_i\mid \mid \vec{S}_j \mid}
\, .  \end{equation}

For comparison, we have also calculated the single
 exchange-integral $j'$
 for the conventional \newline (1-$Co/$n-$Cu$)$_\infty$ superlattices
with the same program, also in scalar-relativistic version, obtained
from the total energy difference $(E^{\uparrow
,\downarrow}-E^{\uparrow ,\uparrow})/(4A)$ analogously to
eqn.~(\ref{eqn1}). 

The results of our study are presented in Fig.~1
and Fig.~2.  For the novel modulated structure it can be seen that the
computed couplings $j$ and $\mid J \mid$ oscillate with the $Cu_n$
spacer thickness in a similar way. The oscillation of $J$ shows that
the system as a whole, rather than the short spacing between the
corresponding $Co$ layers alone, determines the coupling. This is in
agreement with recent experiments of deVries {\it et al.},
\cite{l:vries}, however it is not our main point:
 More important is that the oscillations of the strong coupling $J$
have a large negative bias (i.e.~they favour antiparallel ordering of
the three $Co$ layers of type $Co^{(1)}$ and $Co^{(2)}$; the absolute
value $|J|$ is plotted!) and typically have a much higher amplitude
than the oscillations of the weak coupling $j$. Furthermore it is
remarkable that the weak coupling $j$ remains positive
(i.e.~ferromagnetic) in the range considered, i.e.~for $3\le n \le
6$. This behaviour is in contrast to the behaviour of the exchange
$j'$ in Fig.2, which strongly oscillates from positive to negative
values for the range of $n$-values considered and has negative values
- corresponding to the first antiferromagnetic maximum - in a range
where $j$ is still positive. Of course, we cannot exclude that for
$n\ge 7$ also $j$ could become negative, which might be welcome for
applications, and could show oscillations with $n$ with
similar periods as those of $j'$. We can also not exclude that without
our periodic boundary conditions, the weak exchange $j$ may be
negative for $n \le 6$. All this would not change the conclusions
from our study (see below).

 In any case, for $n=5$ and $6$, see Fig.2, by comparison
with Fig.1 we find that additionally
 the typical magnitude of $j$ (Fig.1) is
significantly smaller than that of $j'$ (Fig.2).
 I.e.~in contrast to $j'$, the coupling $j$ has the correct order of
magnitude (except of the case $n=4$, see below) when compared with
experimental results on similar systems (e.g. \cite{l:blo2}). We
stress this result, since hitherto {\it ab-initio}-calculations of the
present type usually gave by an order of magnitude larger amplitudes
than the experiments (see \cite{l:kr2,l:kr1}). As already mentioned,
the point with $n=4$ is an exception, but even in that case $j$ is
still by a factor 0.59 smaller then $|J|$. The reason for the peculiar
behaviour at $n=4$ may be some kind of reflection- or
confinement-resonance, by which the antiferromagnetic state (ii) from
above is more {\it dis}favoured than for $n=3,5,$ and $6$, compared
with the ferrimagnetic arrangement (i). In any case one should note
that $j(n)$ is ferromagnetic for $n\le 6$ at least, whereas the
corresponding quantity $j'(n)$ in Fig.2 would be antiferromagnetic
(i.e.~$< 0$) for $n\le 4$, and for $n=5$ would have positive values
much larger than $|j(n)|$ for any $n$. Further, in Fig.~2, $j'(n)$
oscillates clearly with a period of $\Delta n \approx 5$, whereas in
Fig.1 the strong coupling $J$ seems to oscillate with a period of only
$\Delta n \approx 2$. In contrast, for the weak coupling $j$ in
Fig.~1, a period cannot be deduced from our data.

Thus we have found that the exchange-biasing slab $Cu_2Co^{(1)}Cu_2$
influences the coupling $j$ (between $Co^{(2)}$ and $Co^{(3)}$ via
$Cu_n$) in an essential way and reduces it substantially.
 Concerning the accuracy of our calculations it should be stated that
even in the worst case the numerical convergence criteria of our
self-consistency loops are still one order of magnitude better than the
small energy differences of large numbers involved in the evaluation
of $j$ and $j'$, so that these results are reliable.

The $j$- and $\mid J \mid$-curves in Fig.~1 separate various magnetic
phases.  As $j$ never crosses zero for $3\le n\le6$ it means that at
least in this region and for vanishing external magnetic field, the
ground state of the multilayers under consideration is the
ferrimagnetic state (i) $[\downarrow\uparrow\downarrow ,\downarrow]$
(up to spin-rotational equivalence), whereas for $n \ge 7$ we cannot
exclude that the antiferromagnetic state (ii) has the lowest
energy. In any case, the ferromagnetic state (iii) should never be the
ground state of our system. 

Although we did not calculate resistances in the different
configurations, and although again the results should depend on the
system as a whole, we expect the following properties, which might be
interesting for applications: Concerning the states (i) and (ii), as
usual, in the antiferromagnetic state (ii) the resistance of the
multilayers should be sensibly larger, so that systems of the present
kind could be interesting for applications as magneto-resistive
sensors or recording heads, if one can fix the orientation of the
strongly coupled ''biasing Co layers'' $Co^{(1)}$ and $Co^{(2)}$
(e.g.~by magnetoelastic interaction with a substrate), which would
still allow for an easy switching of the weakly coupled $Co^{(3)}$
layers. This magneto-resistive effect between the ferromagnetic and
the antiferromagnetic configuration should be stronger for the {\it
CPP}-geometry (current-perpendicular-to-plane) than for the {\it
CIP} (current in plane) geometry \cite{l:sche}, but significant enough
in both cases.

As already mentioned, the physics of quantum confinement,
\cite{l:edwards}, see also \cite{l:krompiewski}, or Fabry-Perrot-like
multiple electron reflection, \cite{l:bruno}, in ultrathin films and
multilayers of the present kind, is also at the origin of the exciting
coupling effects, we have calculated with our extensive calculations.
 We believe that the present {\it ab initio} results may
yield motivation to perform additional model calculations for
modulated systems with the confinement approach. In this way, one
would hopefully get insight also in the reasons of the peculiar
 ''resonance'' at $n=4$, see above, and would be able to deduce an
oscillation period from examination of the large-$n$
limit. Unfortunately, our own work, which involves the accurate
first-principles calculation of 24 different energies for supercells
with up to 20 atoms, can be hardly extended to larger $n$.

\vglue 0.2 truecm In conclusion, our ab-initio calculation with a
spin-polarized LMTO-ASA method for the possible magnetic
configurations of modulated $CoCu_2/CoCu_n$ superlattices of a novel
type has shown the simultaneous presence of {\it strongly} and {\it
weakly} exchange-biased $Co$ monolayers for $3 \le n \le 6$.
  It has been found that the strong coupling across two $Cu$
monolayers is antiferromagnetic and much larger (namely by one order
of magnitude for $n\ne 4$, and still by a factor of $\sim 1.7$ in the
case of $n=4$) than the coupling across the thicker spacer $Cu_n$ with
$3\le n\le 6$, which is ferromagnetic for these $n$ values.
Furthermore, by explicit comparison with conventional
(1-$Co$/n-$Cu$)$_\infty$ monolayers we have shown that in our modulated
system the system as a whole, and not its separate parts, determines
the properties, and that the novel modulated systems behave differently.
The systems considered might be interesting when trying to obtain
exchange-biased spin-valve systems without antiferromagnets,
 where the spin configuration can be easily switched and at
the same time a sensible magneto-resistive effect can be obtained.
 \vspace{0.5cm}

{\bf Acknowledgements}
\vspace{0.5cm}

We would like to thank Profs. F. Stobiecki and G. Bayreuther for 
valuable discussions.
This work has been carried out under the grant no. 2 P 302 005 07 (SK),
 and 
the bilateral project DFG/PAN 436 POL (UK and SK).
 We also thank the  Pozna\'n,
Munich and Regensburg Computer Centres for computing time.

\newpage

{\Large \bf Figure Captions}
\vspace{1cm}

Fig.1: Exchange interactions $J$ and $j$ for the modulated
($CoCu_2CoCu_2CoCu_nCoCu_n$)$_\infty$
superlattices with strongly and weakly exchange-biased Co
monolayers: The {\it strong} antiferromagnetic exchange coupling ($J<0$;
dotted line) acts between two Co monolayers separated by just two $Cu$
monolayers ($Cu_2$), whereas the {\it weak} ferromagnetic coupling ($j>0$)
dashed line) occurs across $Cu_n$ with $3\le n\le 6$. According to
eqn.~(1), $j$ and $|J|$ are $\propto (E_2-E_1)$ and
$(E_3-E_1)$, respectively, where $E_3$, $E_2$ and $E_1$ refer to the
spin configurations (iii) (= ferromagnetic state, highest in energy),
(ii) (= antiferromagnetic state, second highest in energy for $3\le n
\le 6$) and (i) (=ferrimagnetic state) sketched in the text.

\vskip 0.5 truecm
Fig.2: Exchange interaction $j'$ for the conventional
 (1-$Co$/n-$Cu$)$_\infty$ multilayers. Note the different behaviour of
 $j'$ when compared with $j$ in Fig.1.
 The line joining the calculated points is only a guide to the eye.

%\end{document}
%
\newpage
\input epsf
\begin{figure}[htb]
\epsfxsize=16cm
\epsfbox{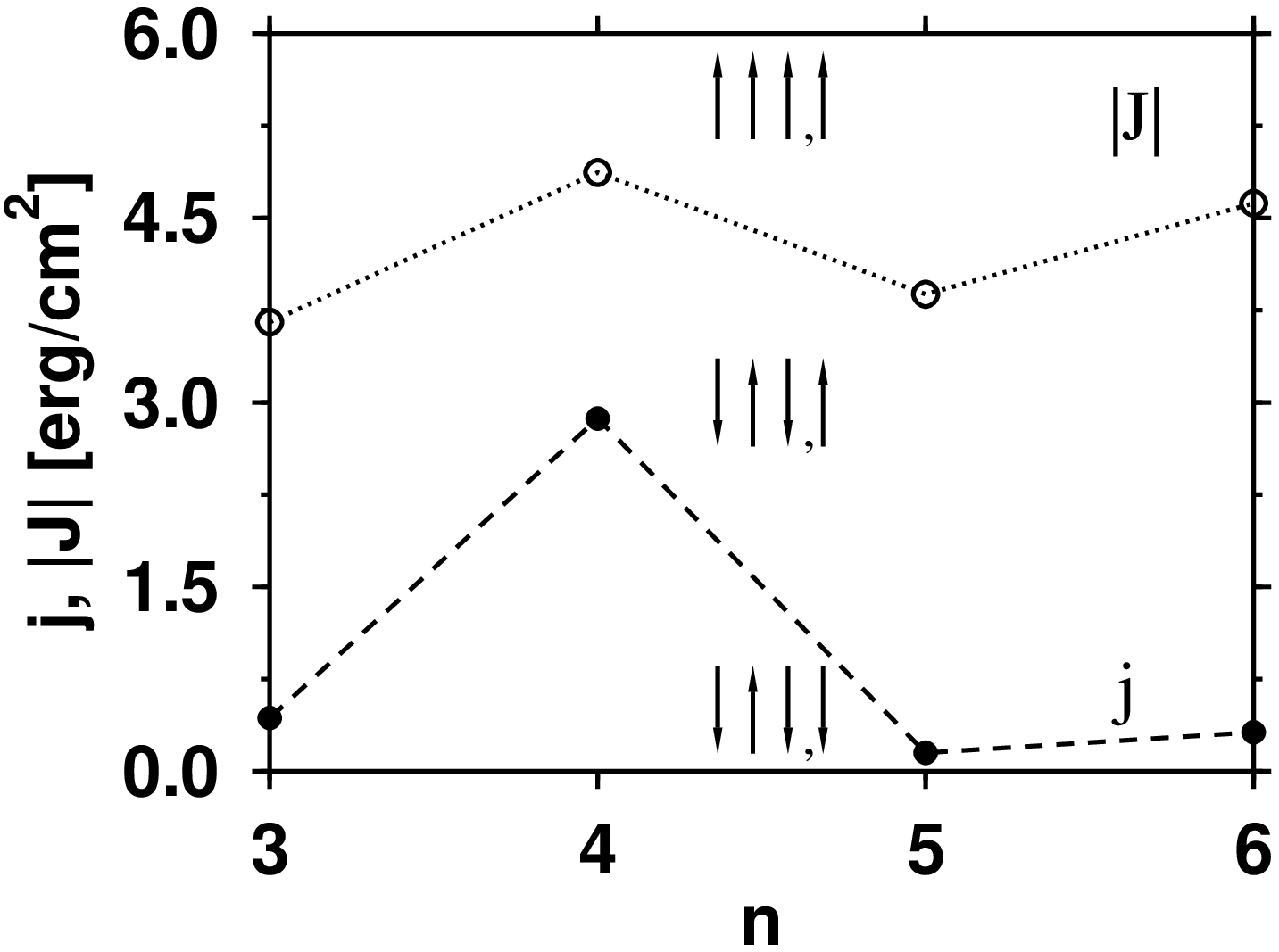}
\end{figure}
\begin{figure}[htb]
\epsfxsize=16cm
\epsfbox{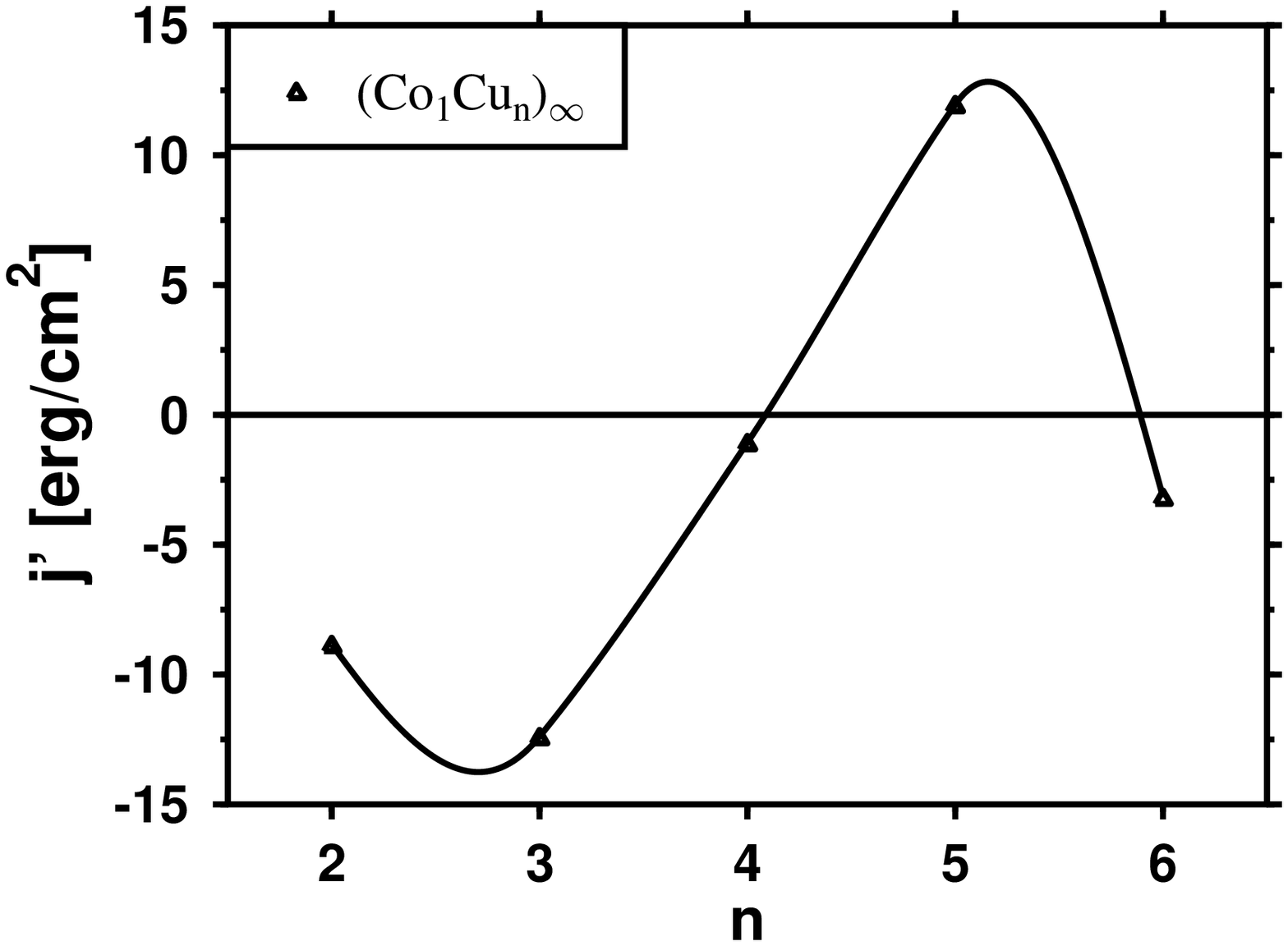}
\end{figure}
\end{document}